\documentclass[aps, prx, 10pt,  amssymb, amsmath, superscriptaddress, tightenlines, twocolumn, notitlepage, longbibliography, citeautoscript] {revtex4-1}
\usepackage{graphicx}
\usepackage{hyperref}
\usepackage{cancel}
\hypersetup{
	unicode=false,          % non-Latin characters in AcrobatÕs bookmarks
	pdftoolbar=false,        % show AcrobatÕs toolbar?
	pdfmenubar=true,        % show AcrobatÕs menu?
	pdffitwindow=false,     % window fit to page when opened
	pdfstartview={FitH},    % fits the width of the page to the window
	pdftitle={},    % title
	pdfauthor={},     % author
	pdfsubject={},   % subject of the document
	pdfcreator={},   % creator of the document
	pdfproducer={}, % producer of the document
	pdfkeywords={}, % list of keywords
	pdfnewwindow=true,      % links in new window
	colorlinks=true,       % false: boxed links; true: colored links
	linkcolor=bleu,          % color of internal links (change box color with linkbordercolor)
	citecolor=bred,        % color of links to bibliography
	filecolor=magenta,      % color of file links
	urlcolor=bleu,           % color of external links
	breaklinks=true
}

\usepackage{amssymb,amsmath,amsfonts}
\usepackage{wasysym}
\usepackage{bm}
\usepackage{physics}
\usepackage{blkarray}
\usepackage[export]{adjustbox}
\usepackage{latexsym}
\usepackage{eucal}
\usepackage{float}
\usepackage{verbatim}
\usepackage[normalem]{ulem}
\usepackage{color}
\usepackage[dvipsnames]{xcolor}
\usepackage{braket}
\usepackage{mathrsfs}
\definecolor{bleu}{rgb}{0.0, 0.5, 0.69}
\definecolor{bred}{rgb}{0.8, 0.25, 0.33}
\definecolor{cnut}{rgb}{0.73, 0.31, 0.28}
\definecolor{gren}{rgb}{0.67, 0.784, 0.2157}
\definecolor{dorange}{RGB}{203,96,21}

\def\i{\mathrm i}
\def\sz{\sigma^{\rm z}}

\def\mx{\mu^{\rm x}}
\def\my{\mu^{\rm y}}
\def\mp{\mu^{+}}
\def\mm{\mu^{-}}

\begin{document}

\title{$U(1)$ symmetry-enriched toric code}

\author{Kai-Hsin Wu}
\email{khwu@bu.edu}
\affiliation{Department of Physics, Boston University, 590 Commonwealth Avenue, Boston, Massachusetts 02215, USA}

\author{Alexey Khudorozhkov}
%\email{alehud@bu.edu}
\affiliation{Department of Physics, Boston University, 590 Commonwealth Avenue, Boston, Massachusetts 02215, USA}

\author{Guilherme Delfino}
%\email{defino@bu.edu}
\affiliation{Department of Physics, Boston University, 590 Commonwealth Avenue, Boston, Massachusetts 02215, USA}

\author{Dmitry Green}
%\email{dmitry.green@aya.yale.edu}
\affiliation{AppliedTQC.com, ResearchPULSE LLC, New York, NY 10065, USA}
\affiliation{Department of Physics, Boston University, 590 Commonwealth Avenue, Boston, Massachusetts 02215, USA}

\author{Claudio Chamon}
%\email{chamon@bu.edu}
\affiliation{Department of Physics, Boston University, 590 Commonwealth Avenue, Boston, Massachusetts 02215, USA}

\date{\today}

\begin{abstract}
  We propose and study a generalization of Kitaev's $\mathbb Z_2$
  toric code on a square lattice with an additional global $U(1)$
  symmetry. Using Quantum Monte Carlo simulation, we find strong
  evidence for a topologically ordered ground state manifold with
  indications of UV/IR mixing, i.e., the topological degeneracy of the
  ground state depends on the microscopic details of the lattice.
  Specifically, the ground state degeneracy depends on the lattice
  tilt relative to the directions of the torus cycles. In particular,
  we observe that while the usual compactification along the
  vertical/horizontal lines of the square lattice shows a two-fold
  ground state degeneracy, compactifying the lattice at $45^\circ$
  leads to a three-fold degeneracy. In addition to its unusual
  topological properties, this system also exhibits Hilbert space
  fragmentation. Finally, we propose a candidate experimental
  realization of the model in an array of superconducting quantum
  wires.
\end{abstract}

\maketitle

\section{Introduction}
Topological phases of matter have been, increasingly, drawing attention from physics and quantum information communities over the past several decades~\cite{wen2013topological}. These special quantum phases exhibit exotic properties such as their topological ground state degeneracy (TGSD) and the excitation content that cannot be captured by conventional Ginzburg-Landau theory. 

A gapped quantum system is considered to possess (\emph{intrinsic}) \emph{topological order} if it has degenerate ground states that are locally indistinguishable. %(i.e., there is no local order parameter)
The TGSD is robust against any local perturbation and depends only on the topology of the underlying spatial manifold \cite{wen2013topological, wegner1971duality, wen1989vacuum, wen1990topological, wen1990ground}. Topologically ordered phases exhibit long-range entanglement %(i.e., they cannot be transformed into a product state through local unitary transformations) 
and support local excitations with fractional exchange statistics (anyons) that cannot be created in isolation by a local process~\cite{arovas1985statistical, wilczek1992disassembling, kitaev2003fault, kitaev2006anyons, stern2008anyons}. Topological order in two spatial dimensions has been extensively studied in various different realizations such as: quantum spin liquids 
%(QSL)
~\cite{read1989statistics, read1991large, wen1991mean, balents1999dual, senthil2000z2, moessner2001resonating, moessner2001short, balents2002fractionalization, kitaev2003fault, levin2005string, savary2017quantum, broholm2020quantum}, fractional quantum Hall 
%(FQH)
 states~\cite{tsui1982two, laughlin1983anomalous, zhang1989effective, stern2008anyons}, superconductors ~\cite{hansson2004superconductors}, topological quantum field theories %(TQFT)
 ~\cite{witten1988tqft, witten1988jones}, etc.

More recently, a notion of \emph{fracton topological order} has been introduced~\cite{chamon2005quantum, bravyi2011topological, haah2011local, yoshida2013exotic, haah2013commuting, vijay2015a, vijay2016fracton, pretko2020fracton}. Fractonic phases of matter exhibit fractionalized excitations (fractons) that cannot be created in pairs. Unlike an anyon, which is created at the end points of a string-like operator (Wilson line) and can freely move across space, a single fracton is immobile, since fractons are created at the corners of membrane- or fractal-like operators. Another difference with the usual topological order is that in fractonic systems the ground state degeneracy (GSD) depends not only on the topology of the manifold, but also on the microscopic properties of the model, such as the system size and the lattice geometry. This is a manifestation of UV/IR mixing in quantum field theory %(QFT)
 ~\cite{seiberg2021twod, seiberg2020threed}.

Additional classification of topological phases of matter arise from an interplay between symmetry and topological order. Even in the absence of intrinsic topological order, a system invariant 
under a symmetry can exhibit several distinct phases that cannot be adiabatically connected to each other, unless one violates the symmetry or closes the energy gap. These phases do not break the considered symmetry and are
known as \emph{symmetry-protected topological} (SPT) phases, with notable examples including free-fermionic topological insulators in 2D \cite{kane2005quantum, kane2005z2, bernevig2006quantum} and 3D \cite{moore2007topological, fu2007topological, roy2009topological, ando2013topological}, topological superconductors \cite{sato2017topological}, as well as interacting bosonic SPTs \cite{su1979solitons, haldane1983continuum, affleck1987rigorous, chen2012symmetry}.
Furthermore, the distinct phases protected by a symmetry need not lack intrinsic topological order. In fact, the presence of symmetry can give rise to adiabatically disconnected phases with distinct topological orders, so called 
\emph{symmetry-enriched topological} (SET) phases, where anyons transform non-trivially under the symmetry \cite{wen2002quantum, levin2012classification, essin2013classifying, hung2013quantized, mesaros2013classification, huang2014detection, lu2016classification, cheng2017exactly, wang2022exactly}.

Despite numerous attempts \cite{levin2005string, kitaev2009periodic, levin2012classification, chen2012symmetry, essin2013classifying, mesaros2013classification, chen2015anomalous, tarantino2016symmetry, lu2016classification, cheng2017exactly, barkeshli2019symmetry, wang2022exactly}, to-date there is no unifying theory for all topological orders. In the case of SETs, many classification attempts rely on constructing representative exactly solvable models that serve as fixed points for each phases. In addition, the majority of the work has been done on finite symmetry groups and only recently has there been an attempt to classify $U(1)$-symmetric SETs \cite{wang2022exactly}.

In this work we pose the question of what happens as one tries to
enrich the usual $\mathbb Z_2$ toric code with a global $U(1)$
symmetry. Although topological order has been previously enriched with
$U(1)$ symmetry through the addition of extra degrees of freedom
\cite{Levin2011, wang2022exactly}, here we take a different approach
and restrict the usual toric code Hamiltonian to only interactions
that are $U(1)$ symmetric. The presence of the extra symmetry imposes
additional constraints on the ground state loop dynamics, compared to
the conventional $\mathbb{Z}_2$ toric code. We refer to the 2D lattice
model as the ``$ U(1)$ symmetry-enriched toric code'' -- or ``$U(1)$
toric code'' for short. Its Hamiltonian is not a sum of commuting
projectors, and hence not exactly solvable in any obvious
way. However, the model does not have a sign problem, which allows us
to study the model via large scale Quantum Monte Carlo (QMC) simulation.

One of the motivations to study the model is its relation to the so
called WXY model \cite{chamon2021superconducting}, which consists of
two-body interaction terms between ``matter'' and ``gauge'' spins,
located at the vertices and edges of a square lattice,
respectively. It possesses combinatorial $\mathbb{Z}_2$ gauge symmetry
\cite{Chamon2020, chamon2021superconducting, Wu2021,
  green2022constructing, yu2022abelian} in addition to a global $U(1)$
symmetry.  The $U(1)$ toric code is believed to emerge after one
integrates out the matter spins in the WXY model.  However, unlike the
former, the WXY model is difficult to study numerically since the unit
cell is too large for exact diagonalization and the sign problem
prevents us from employing Quantum Monte Carlo.

Through our numerical studies on the $U(1)$ toric code, we observe a
degenerate ground state manifold where non-local quantum numbers are
necessary to distinguish the states.  We also observe the existence of
UV/IR mixing, which manifests itself in the change of the ground state
degeneracy upon contracting the lattice on a torus through different
compactification vectors.  A similar flavor of a UV/IR mixing was
explored in Ref. \cite{Rudelius2021}. It differs from the usual
manifestation of UV/IR mixing in fractonic models, where the GSD
typically depends only on the system size. In addition to its unusual
GSD properties, the model displays Hilbert space
fragmentation~\cite{khemani2020localization, sala2020ergodicity,
  moudgalya2022hilbert}. Finally, we propose a physical realization of
the model in a mesh of Josephson-coupled superconducting quantum
wires.

\section{Lattice Model}
\label{sec:model}
One simple way to arrive at the model of interest is to begin with Kitaev's $ \mathbb{Z}_2 $ toric code~\cite{kitaev2003fault} and then enrich it by an additional global $U(1)$ symmetry. (A second way, less direct but closer to experimental settings, is presented in Sec.\ref{sec:realization}.) 
Consider a square lattice with sites $s\equiv (i,j)$, where $i,j$ are the $x$- and $y$-coordinates, and elementary lattice vectors  
$\hat{x} \equiv 2\hat{e}_x = (1,0), \hat{y} = 2\hat{e}_y = (0,1)$.
A spin-1/2 degree of freedom is located on each link $\ell$. We say $\ell \in s$ if the link $\ell$ is one of the four links adjacent to the site $s$, i.e., if $\ell = s \pm \hat{e}_{x,y}$. We define a star operator ${A}_s(\theta)$ as the product of four spin-1/2 operators $\sigma_\ell^{\theta} =\cos(\theta)\,\sigma^x_{\ell} + \sin(\theta)\,\sigma^y_{\ell}$ on the links adjacent to site $s$,
\begin{align}
	\begin{split}
		{A}_s(\theta) = \prod_{\ell \in s} \sigma_\ell^\theta.
	\end{split}
\end{align}  
$\sigma^x$, $\sigma^y$, and $\sigma^z$ are the usual Pauli spin operators. The generalized star operator ${A}_s(\theta)$ for any angle $\theta$ is invariant under a $ \mathbb{Z}_2 $ local transformation, generated by the product of
$\sigma^z$ operators on the links of a given plaquette $p$:
\begin{align}
	\begin{split}
		B_p = \prod_{\ell\in \partial p} \sigma^z_\ell\;.
	\end{split}
\label{z2}
\end{align}
In other words, $[{A}_s(\theta), B_p]=0$, for all $s$, $p$ and $\theta$, which is similar to the usual toric code for fixed $\theta$.

The global $U(1)$ symmetry can be introduced by
averaging ${A}_s(\theta)$ over all angles:
\begin{align}
	\begin{split}
		\mathcal{A}_s = \dfrac{1}{2\pi} \int_0^{2\pi}d\theta\;{A}_s(\theta)\;.
	\end{split}
\label{star}
\end{align}
Keeping only the surviving terms in the integral, this star operator can be written in terms of spin raising and lowering operators as  
\begin{align}
	\begin{split}
		\mathcal{A}_s & = \sigma^+_{s+\hat{e}_x}\sigma^+_{s+\hat{e}_y}\sigma^-_{s-\hat{e}_x}\sigma^-_{s-\hat{e}_y}
		\\
		& + \sigma^+_{s+\hat{e}_x}\sigma^-_{s+\hat{e}_y}\sigma^+_{s-\hat{e}_x}\sigma^-_{s-\hat{e}_y}
		\\
		& + \sigma^+_{s+\hat{e}_x}\sigma^-_{s+\hat{e}_y}\sigma^-_{s-\hat{e}_x}\sigma^+_{s-\hat{e}_y} + \text{h.c.},
	\end{split}
\label{eq.A_as_pm}
\end{align}
with $ \sigma^{\pm}_\ell  = \left( \sigma_\ell^x \pm i\sigma^y_\ell \right)/2$. It is easy to check that the stars $\mathcal{A}_s $ in Eq.~\eqref{star} are invariant under a global $z$-axis rotation,
\begin{align}
	\begin{split}
		U_z = \exp\left (-i\dfrac{\alpha}{2} M_z\right ),
	\end{split}
\label{u1}
\end{align}
where $ M_z = \sum_\ell \sigma^z_\ell $  is the total magnetization in the $z$-direction. Namely
\begin{align}
	\begin{split}
		U_z \;\mathcal{A}_s \;U_z^\dagger = \dfrac{1}{2\pi} \int_0^{2\pi} d\theta\; {A}_s(\theta+\alpha) = \mathcal{A}_s.
	\end{split}
\end{align}
The magnetization conservation is also apparent from Eq.~\eqref{eq.A_as_pm}, since, in the $ \sigma^z $ basis, every term flips exactly two spins up and two spins down.
The Hamiltonian can now be formally written as the sum over all possible star and plaquette operators,
\begin{align}
	\begin{split}
		H = -\lambda_{\mathcal A}\;\sum_s \mathcal{A}_s - \lambda_B\;\sum_p B_p.
	\end{split}
\label{eq:Hamiltonian}
\end{align}
We note that this model is not a sum of commuting projectors: neighboring stars do not commute, i.e., $ [\mathcal{A}_s, \mathcal{A}_{s'}] \neq 0$ if $s$ is adjacent to $s'$.

One can easily check that the Hamiltonian commutes with total magnetization $M_z$ and all the plaquettes $\{B_p\}$,
\begin{align}
	\begin{split}
		[H, M_z]=0\quad \text{and} \quad [H, B_p]=0 \quad \forall p, 
	\end{split}
\end{align}
which are conserved quantities associated to the global $U(1)$ and local $\mathbb Z_2$ gauge symmetry, respectively. 

For any local closed loop $ \gamma $ composed of a sequence of connected links, the loop operator
\begin{align}
	\begin{split}
		W(\gamma) = \prod_{\ell\in \gamma} \sigma^z_\ell
	\end{split}
\end{align}
commutes with the Hamiltonian, $ [H, W(\gamma)] =0$, and can be represented as a product of plaquette operators $B_p$ enclosed by $\gamma$.
When putting the system on a torus, two additional loop-operators, $ W_x $ and $ W_y$, defined along the shortest non-contractible loops that wind around the torus in the $ x $ and $ y $ directions are also conserved.
The $U(1)$ toric code can be block-diagonalized in the common eigen basis of these operators, so that each sector is characterized by a set of independent conserved quantities $\{M_z,W_x,W_y,\{B_p\}\}$.
We are particularly interested in the four sectors categorized by $W_x = \pm 1, W_y=\pm 1$, as they underscore the topological features. In the following, we refer to these four sectors as topological sectors, with respective quantum numbers $(W_x,W_y) = (+,+),( +,-), (-,+)$ and $(-,-)$. 

Interestingly, we find that within each symmetry block the Hilbert
space further splits into multiple even smaller fragments (or Krylov
subsectors), indicating that the system exhibits Hilbert space
fragmentation \cite{khemani2020localization, sala2020ergodicity,
  moudgalya2022hilbert}. An exact diagonalization (ED) study that we
carried out suggests that in the vicinity of zero magnetization, the
Hilbert space of each symmetry sector is dominated by a single large
fragment, an indication of weak Hilbert space fragmentation. This
implies that in these symmetry sectors the lowest energy eigenstates
belong to this large fragment (see Appendix~\ref{App.Frag}).  We note
that the fragmentation structure for a different system whose dynamics
we can map to that of the $U(1)$ toric code in the sector of
$\{W_x=+1,W_y=+1, \{B_p=+1\}\}$ has been shown to be
weak~\cite{hart2022hilbert,balducci2022localization,balducci2022interface,yoshinaga2022emergence},
which is consistent with our ED results.

\section{Numerical results}	

The $U(1)$ toric code in Eq.~(\ref{eq:Hamiltonian}) is free of a sign
problem, which allows us to study it via the Stochastic Series
Expansion Quantum Monte Carlo (SSE QMC) method using a generalized
version of the sweeping cluster update
algorithm~\cite{SandvikSSE2003,sandvik2010computational,yan2019sweeping}.
This algorithm is designed to perform simulations not only within any
symmetry sector, but within any Hilbert space fragment, specified by
an initial basis state from that fragment (See Appendix~\ref{App.SCA}
for technical details). In our studies we always choose an initial
state within the largest fragment in the symmetry sector of interest.

In this work, we focus on the even parity sector where all $B_p = +1$,
which encompasses the ground states of the Hamiltonian
\eqref{eq:Hamiltonian} for sufficiently large $\lambda_B$ (we set
$\lambda_\mathcal{A}=1$ henceforth). Within the even parity sector, we
then study all four topological sectors that are characterized by
$W_x = \pm 1$ and $ W_y = \pm 1$. We are particularly interested in
whether the lowest energy states from different topological sectors
are degenerate, as this degeneracy might indicate the existence of
topological order in the system.

\begin{figure}[t]
\includegraphics[scale=0.9]{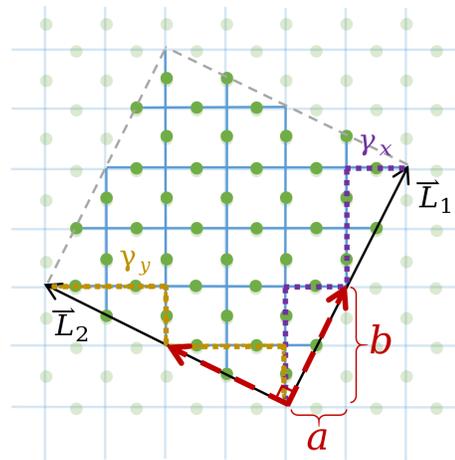}
\caption{An example of a lattice with compactification $a=1, b=2$ and
  linear size $L=2$. Any vector $\vec{r}$ is identified with vectors
  $\vec{r} + \vec{L}_1$ and $\vec{r} + \vec{L}_2$. The whole lattice
  is shown in bright colors, while the shaded region denotes repeating
  parts of the lattice due to the periodic boundary condition. Two non-contractible loops $\gamma_x$ and $\gamma_y$ shown as purple and yellow dotted lines respectively along two compactification vectors $\vec{L}_1$ and $\vec{L}_2$. }
\label{Fig.geocmptfy}
\end{figure}

In addition, we examine the dependence of the ground state properties
of the lattice on a torus with various compactifications.  We
introduce two orthogonal compactification vectors $\vec{L}_1$ and
$\vec{L}_2$, parameterized by two non-negative integers $a$ and $b$,
as
\begin{align}
	\begin{split}
		\vec{L}_1 &= L(a \hat{x} + b \hat{y})\\
		\vec{L}_2 &= L(-b \hat{x} + a \hat{y}),
	\end{split}
\end{align}	
where $L$ (a positive integer) is the linear system size. We choose
the vector $(a,b)$ to be the shortest integer vector in its direction,
i.e., $a$ and $b$ are coprimes.  Vectors $\vec{L}_1, \vec{L}_2$ define
the compactification scheme in the sense that any spatial vector
$\vec r$ is identified with vectors $\vec{r} + \vec{L}_1$ and
$\vec{r} + \vec{L}_2$. An example of a small lattice with a
non-trivial compactification scheme is shown in
Fig.\,\ref{Fig.geocmptfy}.
%%%%%%%%%%%%%%%%%%%%%%%%%%%%%%%%%%%%%%%%%%%%%%%%%%%%%%%%%%%%%%%%%%%%%%%
%%%%%%%%%%%%%%%%%%%%%%%%%%%%%%%%%%%%%%%%%%%%%%%%%%%%%%%%%%%%%%%%%%%%%%%
%We define the non-contractible loop operators as products of spin
%operators along the $\vec{L}_1$ and $\vec{L}_2$ directions. More
%specifically, we define the unit vectors
%$\vec{\ell}_1 = \vec L_1/ |\vec L_1|$ and
%$\vec{\ell}_2 = \vec L_2/ |\vec L_2|$ and the string sequences of
%points
%$\gamma_x = \{\vec{\ell}_1, 2\vec{\ell}_1, \ldots, |\vec L_1|
%\vec{\ell}_1\}$ and
%$\gamma_y = \{\vec{\ell}_2, 2\vec{\ell}_2, \ldots, |\vec L_2|
%\vec{\ell}_2\}$.  
%The loop operators are given as
%\begin{align}
%W_x &= \prod_{i\in \gamma_x} \sigma^{z}_{i} \\
%W_y &= \prod_{i\in \gamma_y} \sigma^{z}_{i} .
%\end{align}

We define the non-contractible loops $\gamma_x$ and $\gamma_y$ along $\vec{L}_1$ and $\vec{L}_2$ directions as shown in Fig.~\ref{Fig.geocmptfy} by the purple and yellow dotted lines, respectively. 
The non-contractible loop operators are given by the products of spin operators along these two loops as
\begin{align}
W_x &= \prod_{i\in \gamma_x} \sigma^{z}_{i}, \\
W_y &= \prod_{i\in \gamma_y} \sigma^{z}_{i} .
\end{align}

Note that there is freedom in the choice of non-contractible strings
entering the definition of the operators $W$, since two strings
operators going around the torus in the same direction can be deformed
into each other by multiplications of $B_p$ operators.

Also note that even though the symmetry generators,
$M_z, W_x, W_y, \{B_p\}$, are independent of each other, their quantum
numbers might be incompatible. Compatibility of quantum numbers
depends on the compactification scheme, as well as the linear size
$L$. For example, in the case of $a=1$, $b=0$ compactification, it is
not possible to have a zero magnetization state within the $(+,+)$ and
$(-,-)$ topological sectors when the linear size $L$ is odd (see
Appendix~\ref{App.Mag}).

Below we explore in detail two lattice compactifications on a torus:
A) $0^\circ$-tilt compactification, corresponding to $a=1$, $b=0$ --
the usual compactification along the vertical and horizontal lines of
the square lattice; and B) $45^\circ$-tilt compactification, with
$a=1$, $b=1$. We focus on the cases with $L$ even, for which we
observe that the ground states have magnetization $M_z = 0$ in of all
topological sectors. Cases with $L$ odd are explored in
Appendix~\ref{App.odd}, and compactifications with other tilt angles
are discussed in Appendix~\ref{App.GenCptfy}.

\subsection{$0^\circ$-tilt compactification}

\begin{figure}[ht]
  \includegraphics[width=\linewidth]{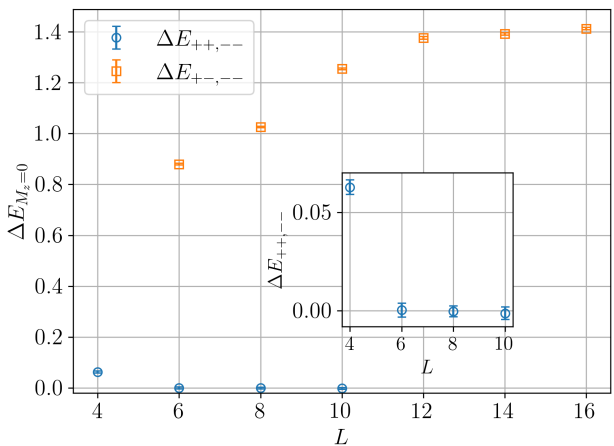}
  \caption{Energy gaps between the ground states in different
    topological sectors as a function of system size $L$ in the
    $0^\circ$-tilt
    compactification. $E_{W_x\,W_y,W^\prime_x\,W^\prime_y}$ labels the
    energy difference between sectors $(W_x,W_y)$ and
    $(W^\prime_x,W^\prime_y)$. For all system sizes, the sector
    $(-,-)$ has the lowest energy. The inset shows that the finite
    size gap $\Delta E_{++,--}\to 0$ as $L\rightarrow \infty$. The
    results indicate that the system has a two-fold TGSD.}
  \label{Fig.0m0gap}
\end{figure}

We observe a finite energy gap of order ${\cal O}(1)$ between sectors
$(W_x, W_y) = (+,-)$ and $(-,-)$ for all system sizes $L$
considered. The sectors $(+,-)$ and $(-,+)$ have identical energy
spectra due to the $C_4$ rotation symmetry of the lattice. On the
other hand, the energy difference between sectors $(+,+)$ and $(-,-)$
vanishes as $L$ increases, with the lowest energy state in the sector
$(-,-)$ in all cases (see Fig.~\ref{Fig.0m0gap}). We find that this
energy difference becomes essentially zero within error bars beyond
system sizes as small as $L=6$. We conclude that the system has a
two-fold degeneracy associated to sectors $(+,+)$ and $(-,-)$, to
which we henceforth refer as a topological degeneracy (TGSD)
distinguished by quantum numbers associated to operators $(W_x, W_y)$
defined on non-contractible loops.

%%%%%%%%%%%%%%%%%%%%%%%%%%%%%%%%%%%%%%%%%%%%%%%%%%%%%%%%%%%%%%%%%%%%%%%
%%%%%%%%%%%%%%%%%%%%%%%%%%%%%%%%%%%%%%%%%%%%%%%%%%%%%%%%%%%%%%%%%%%%%%%

The observation of an ${\cal O}(1)$ energy separation between topological sectors suggests that the system is gapped. However, this observation alone does not rule out the possibility that the system is gapless within each of the $(+,+)$ and $(-,-)$ ground state sectors. 
To establish that the system is, in fact, gapped within each of these sectors, we compute the spin-spin correlation functions along the $x$-direction, $C(r)$, and along the $45^\circ$ diagonal direction, $C_d(r)$, defined as 
\begin{subequations}
  \label{eq.Corr}
  \begin{align}
  C(r) &=  \frac{1}{L} \sum_\alpha \sigma^z_{(\hat{e}_x + \alpha\hat{y})} \sigma^z_{(\hat{e}_x + \alpha\hat{y}) + r \hat{x} }, \\
  C_d(r) &= \frac{1}{L} \sum_\alpha \sigma^z_{(\hat{e}_x + \alpha \hat{e}_{\bar{x}y} )} \sigma^z_{(\hat{e}_x + \alpha \hat{e}_{\bar{x}y}) + r \hat{e}_{xy}},
  \end{align}
\end{subequations}
where $\hat{e}_{xy} = \hat{e}_x + \hat{e}_y$, $\hat{e}_{\bar{x}y} = -\hat{e}_x + \hat{e}_y$.
We observe that both correlation functions decay rapidly to zero within a short distance on the order of two lattice sites, as shown in Fig.~\ref{Fig.0degCorr}(a), for the $(-,-)$ sector (see Appendix~\ref{App.Corr} for correlations in other topological sectors). This rapid decay is consistent with the absence of long-range magnetic order and
%. The rapid decay of the spin-spin correlation function 
provides evidence that the system is gapped within each topological sector.

%%%%%%%%%%%%%%%%%%%%%%%%%%%%%%%%%%%%%%%%%%%%%%%%%%%%%%%%%%%%%%%%%%%%%%% 
\begin{figure}[h]
  \includegraphics[width=\linewidth]{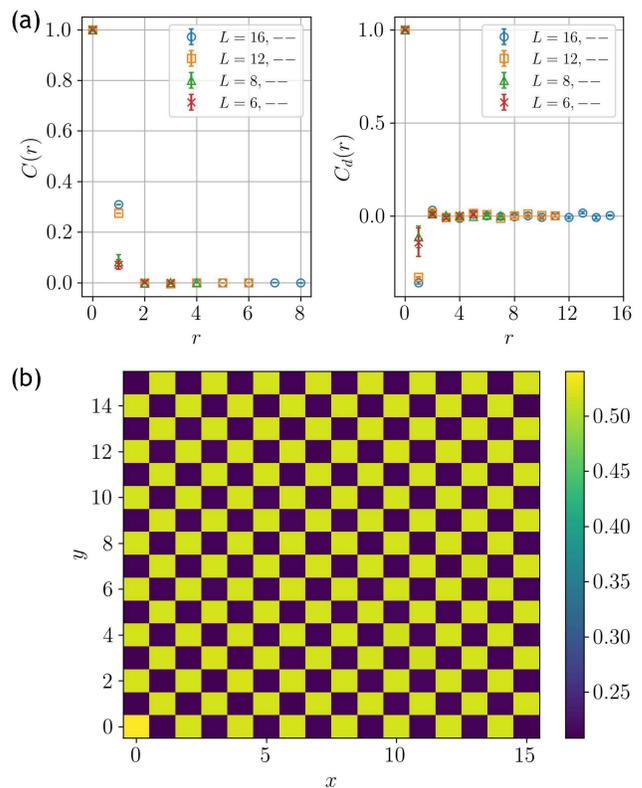}
  \caption{(a) Spin-spin correlation function in the $x$- and
    $45^\circ$-direction ($C(r)$ and $C_d(r)$ respectively) for
    different system sizes, $L$. The correlation function decays to
    zero rapidly, indicating that the system is gapped. (b) Intensity
    plot of the star-star correlation
    $\left<\mathcal{A}_{s=(0,0)}\;\mathcal{A}_{s'=(x,y)}\right>$,
    indicating translational symmetry breaking. The system size is
    $L=16$. Both results shown here correspond to the $0^\circ$-tilt
    compactification, in the sector $(-,-)$ and with $M_z=0$. }
  \label{Fig.0degCorr}
\end{figure}
%%%%%%%%%%%%%%%%%%%%%%%%%%%%%%%%%%%%%%%%%%%%%%%%%%%%%%%%%%%%%%%%%%%%%%%

We observe a spatial checkerboard pattern in the measurements of the
star-star correlator $\left<\mathcal{A}_s\;\mathcal{A}_{s'}\right>$,
shown in Fig.~\ref{Fig.0degCorr}(b). This pattern suggests that the
ground states spontaneously break translation symmetry. This staggered
pattern appears in both degenerate topological sectors $(-,-)$ and
$(+,+)$, indicating the coexistence of spontaneous symmetry breaking
with topological degeneracy, i.e., the total ground state degeneracy
is 4, the product of the two-fold TGSD by a factor of 2 originating
from the symmetry breaking.

\subsection{$45^{\circ}$-tilt compactification}

\begin{figure}[h]
  \includegraphics[width=\linewidth]{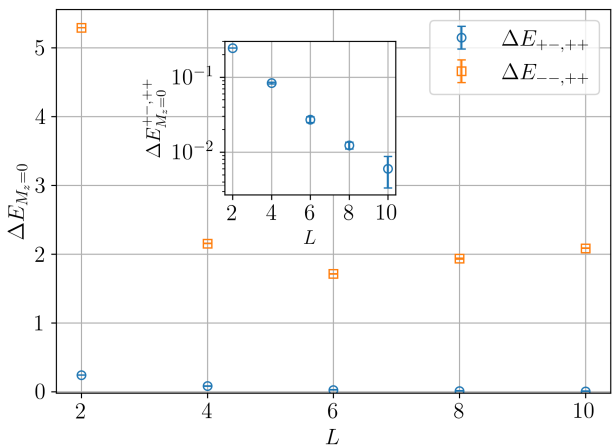}
  \caption{Energy gaps between the ground states in different
    topological sectors as a function of system size in the
    $45^\circ$-tilt compactification. The notation is common to that
    of Fig.~\ref{Fig.0m0gap}. The state in the $(-,-)$ sector has
    higher energy, and is separated from the states in the other three
    sectors $(+,+),(+,-)$, and $(-,+)$, which are degenerate. [Note
    that $(+,-)$ and $(-,+)$ sectors have identical energy spectra due
    to the $C_4$ rotation symmetry of the lattice.]}
  \label{Fig.45FSzgap}
\end{figure}

We repeat the studies above for the $45^{\circ}$-tilted lattice. The
essential observation here is that the state in the $(-,-)$ sector has
higher energy than the states in the other three sectors
$(+,+),(+,-)$, and $(-,+)$, which are degenerate, thus yielding a
three-fold TGSD (see Fig.~\ref{Fig.45FSzgap}). We find, again, that
the spin-spin correlation functions are fast decaying for all three
states in the ground state manifold, just as in the case of
$0^\circ$-tilt compactification. This suggests the absence of long
range magnetic order and provides strong evidence for a gapped ground
state manifold (see Appendix.~\ref{App.Corr}).
		
\section{Discussion of the ground state degeneracy}
The results above establish numerically that the TGSD depends on the
tilt angle of the compactification of the lattice. For the
$0^\circ$-tilt, we find a TGSD of two (sectors $(+,+)$ and $(-,-)$),
while for the $45^\circ$-tilt we find a TGSD of three (sectors
$(+,+), (+,-), $ and $(-,+)$). This result is peculiar for two
reasons. First, a change of the TGSD upon changing the
compactification of the lattice is a clear manifestation of UV/IR
mixing, a feature quite common in gapped fractonic phases. In
fractonic systems the TGSD typically depends on the system size and on
the relative dimensions of the lattice \cite{chamon2005quantum,
  bravyi2011topological, haah2011local,yoshida2013exotic,
  haah2013commuting, vijay2015a,vijay2016fracton}, while here it
depends on the choice of compactification directions (vectors)
defining a torus.

Second, the three-fold topological degeneracy in the $45^\circ$-tilt
poses a puzzle.  The $U(1)$ toric code Hamiltonian
Eq.~\eqref{eq:Hamiltonian} is time-reversal symmetric.  The observed
three-fold topological degeneracy is quite unusual when coming from a
time-reversal ($T$) invariant $ \mathbb Z_2 $ gauge theory. Typically,
topological quantum field theories with ${T}$-symmetry are
characterized by Hilbert space dimensions that are either the square
of an integer or numbers that decompose into Pythagorean prime
ones~\cite{demastro2019arithmetic}. Neither is consistent with a
degeneracy of three. One logical possibility is that the $T$-symmetry
is spontaneously broken. If this is the case, it is not manifest
through long-range magnetic order, since we find that the ground
states have zero magnetization and that the spin-spin correlation
functions decay exponentially.

Since we do not have an analytical solution with which to compare the numerical features that we observe, we follow a phenomenological approach to see what features the theory must have in order to be consistent with our numerical results. Let us posit
the existence of non-local tunneling operators $T_{x,y}$, which are not
necessarily unitary, that flip the eigenvalues of the
non-local winding loop operators $W_{x,y}$. As opposed to the usual toric code, one cannot explicitly write down these operators (or at least we do not know of a way, yet).

\begin{figure}[ht!]
	\includegraphics[width=\linewidth]{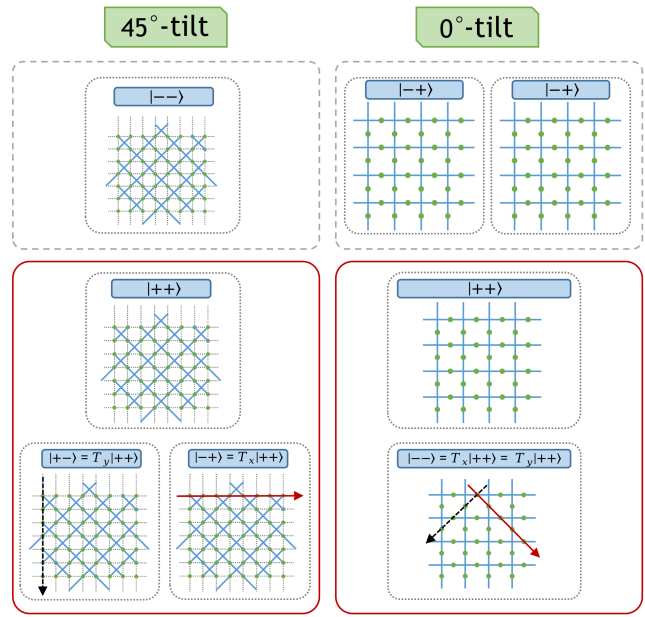}
	\caption{
	%\textcolor{red}{Add $T^{45^\circ}_x$, $T^{45^\circ}_y$ labels (?)}
	 Mapping between topological sectors in $0^\circ$-tilt (right side) and $45^\circ$-tilt (left side) compactifications. The ground state manifold is depicted with the red frames, while topological sectors with excited lowest energy states are depicted with gray dashed frames. We assume existence of non-local and non-unitary tunneling operators, $T^{45^\circ}_x$ (red arrows) and $T^{45^\circ}_y$ (dashed black arrows), that in the $45^\circ$-tilt case take the state $\ket{++}$ to states $\ket{-+}$ and $\ket{+-}$, respectively. Simultaneous application of both tunneling operators to $\ket{++}$ annihilates the state, and therefore $\ket{--}$ does not belong to the ground state manifold. In the $0^\circ$-tilt case, assuming the same orientation of the tunneling operators with respect to the microscopic details of the lattice, both $T^{45^\circ}_x$ and $T^{45^\circ}_y$ take state $\ket{++}$ to $\ket{--}$, and hence states $\ket{+-}$, $\ket{-+}$ remain out of the ground state manifold.}
	\label{fig:cptfy}
\end{figure}

Consider first the case of $45^\circ$-tilt compactification, and
operators $T^{45^\circ}_x$ and $T^{45^\circ}_y$, depicted as a red
solid-arrow and a black dashed-arrow in Fig.~\ref{fig:cptfy}; these
operators change the eigenvalues of $W_{x}$ and $W_y$,
respectively. The three ground states can be indexed as
\begin{align}
	\begin{split}
		\ket{++}\;,
		\;
		\ket{-+}\sim T^{45^\circ}_x\,\ket{++}\;,
		\;
		\ket{+-}\sim T^{45^\circ}_y\,\ket{++}
		\;.
	\end{split}
\end{align}
To be consistent with the numerical result, that the fourth state
$\ket{--}$, is an excited state and not yet another ground state, the
application of the product $T^{45^\circ}_x T^{45^\circ}_y$ to the
reference state $\ket{++}$ must be orthogonal to the ground state
manifold, or equivalently, i.e., it must annihilate the reference
state in the ground state manifold,
\begin{align}
	\begin{split}
		T^{45^\circ}_x\,T^{45^\circ}_y\,\ket{++}\; \sim  0
		\; .
	\end{split}
\label{eq:fourth}
\end{align}
This scenario parallels that of $SU(2)_2$ topological order (hosting
Ising anyons), where one can insert fluxes (corresponding to the
tunneling operators, $T^{45^\circ}_x$ or $T^{45^\circ}_y$) through one
or the other hole of the torus and switch ground states, but not
insert flux through both (see, for example, Ref.~\cite{Oshikawa20071477}). The net effect is to make
 the fourth ground state inaccessible as in
Eq.~\eqref{eq:fourth}. (Refs.~\cite{xavier2022nonabelian,
  iadecola2019ground} give examples of systems with
$SU(2)_2$ topological order where the tunneling operators
$T_{x,y}$ can be constructed and their algebra is studied.)

These tunneling operators also allow us to propose an heuristic
argument that connects the topological ground state degeneracy in the
two compactification schemes. For the case of $0^\circ$-tilt
compactification, the two tunneling operators $T^{45^\circ}_x$ and
$T^{45^\circ}_y$ (again depicted as a red solid-arrow and a black
dashed-arrow in Fig.~\ref{fig:cptfy}) wind across the torus along the
$+45^\circ$ and $-45^\circ$ directions. Along both directions, the
tunneling operators flip both winding loop eigenvalues $W_x$ and
$W_y$, and we write the two ground states as
\begin{align}
	\begin{split}
		\ket{++}\;,
		\;
		\ket{--}\sim T^{45^\circ}_x\,\ket{++}\; \sim T^{45^\circ}_y\,\ket{++}
		\;.
	\end{split}
\end{align}
Because both $T^{45^\circ}_x$ and $T^{45^\circ}_y$ have the same
action on $\ket{++}$, they provide us with only one additional state,
$\ket{--}$, and thus a topological ground state degeneracy of two in the
$0^\circ$-tilt compactification.

Again, the argument for the exclusion of the fourth sector,
$\ket{--}$, from the ground state manifold in the case of
$45^\circ$-tilt compactification parallels that in the case of Ising
anyons. These arguments suggest a logical possibility that the
$U(1)$ toric code may realize non-Abelian topological order.
A thorough investigation of this possibility is left for future work.

One way to test the robustness of the ground state
degeneracy is to perturb the Hamiltonian \eqref{eq:Hamiltonian} using
arbitrary local terms, such as longitudinal and transverse magnetic
fields
\begin{align}
	\begin{split}
		H\rightarrow \tilde H = H-g_x\sum_\ell \sigma^x_\ell -g_z\sum_{\ell} \sigma^z_\ell.
	\end{split}
\end{align}
However, we encounter the following difficulties in numerically
carrying out such tests using both ED or QMC. ED is limited to small
system sizes, posing the numerical challenge of distinguishing effects
due to the perturbation from those due to the finite size (the
correlation length, even if small as suggested by the decay of
spin-spin correlations in Fig.~\ref{Fig.0degCorr}, is not zero). In
the case of QMC, our update algorithm is build so as to conserve all
quantum numbers, $\{B_p\}, M_z, W_x, W_y$, and cannot incorporate
perturbations that break the conservation of these quantities without
significant modification of the update scheme. (It is unclear what
modifications would be required in the QMC update algorithm to
accommodate such perturbations.)

\section{A physical realization of the star-term of the $U(1)$ toric code} 
\label{sec:realization}

\begin{figure}[h]
  \includegraphics[width=\linewidth]{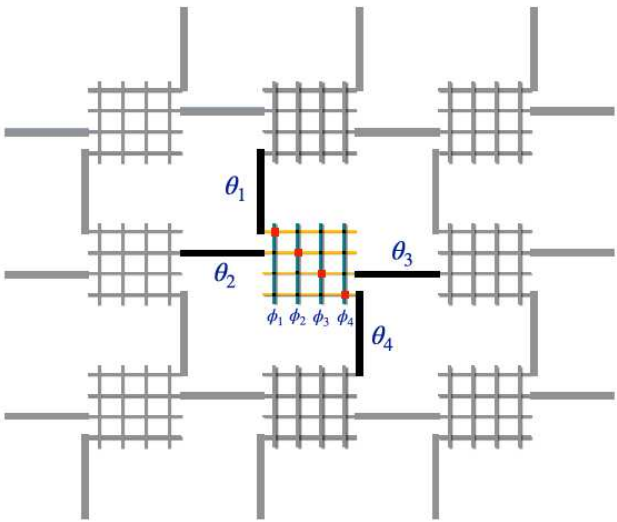}
  \caption{A proposed physical realization of the star term $\mathcal{A}_s$ in the $U(1)$ toric code lattice Hamiltonian. The center ``waffle" is highlighted as an example. It is composed of intersecting superconducting wires coupled by Josephson junctions. The junctions in the diagonal (red) denote $\pi$ couplings. Vertical wires (blue) are ``matter" degrees of freedom labeled by phase $\phi_n$. Horizontal wires (gold) are ``gauge" degrees of freedom with phases $\theta_i$. Only the gauge degrees of freedom (black) couple to other, neighboring, waffles.}
  \label{Fig.JJ_lattice}
\end{figure}

Here we illustrate how the four-spin interaction term $\mathcal{A}_s$
in our Hamiltonian in Eq.~(\ref{eq.A_as_pm}) appears naturally in a
physical set-up proposed in Ref.~\cite{chamon2021superconducting}
using arrays of superconducting quantum wires coupled via Josephson
junctions. Consider, for each star $s$, a $4\times 4$ array of
vertical and horizontal wires intersecting at 16 crossings, as
depicted in Fig.~\ref{Fig.JJ_lattice}. Each of the four vertical wires
$n=1,\ldots,4$ is coupled to each of the four horizontal wires
$i=1,\ldots,4$ by a Josephson junction. The sign of each coupling is
encapsulated by a matrix $W$ with diagonal elements $W_{n=i} = -1$
(corresponding to a $\pi$-junction) and off-diagonal elements
$W_{n\ne i} = +1$ (corresponding to a regular junction).

The Hamiltonian at a given site $s$ for such
a system is given by
\begin{subequations}
  \begin{align}
&
    H = H_J+H_K
    \label{eq:JJ-H}
\end{align}
with
\begin{align}
&
  H_J = -J\; \sum_{n,i} W_{ni}\;\cos(\phi_n-\theta_i)
  \label{eq:JJ-J}
\end{align}
and
\begin{align}
&H_K = 
  \frac{1}{2C_{\text{m}}}\sum_{n}\;q_n^2
  +
  \frac{1}{2C_{\text{g}}}\sum_i\;Q_i^2
  \;,
  \label{eq:JJ-K}
\end{align}
\end{subequations}
where $C_{\text{m}}$ and $C_{\text{g}}$ are the
self-capacitances. $\phi_n$ and $\theta_i$ are the superconducting
phases in each wire and $q_n$ and $Q_i$ are their conjugate charges,
respectively. On the lattice, we refer to the $\phi$ as ``matter''
phases and they are not connected to wires on any other site.  On the
other hand, we refer to the $\theta$ as ``gauge'' phases and they are
shared by neighboring sites. (Here we focus on a single star; on a
lattice, gauge wires are shared between neighboring stars.)

The form of the $W$-matrix guarantees that this Hamiltonian has local
$\mathbb{Z}_2$ symmetry, per combinatorial gauge symmetry (CGS). At the same time, this Hamiltonian
also has global $U(1)$ symmetry, hence it is a natural starting point
for the $U(1)$ toric code.

There are two types of limits that one usually considers in Josephson
junction Hamiltonians -- phase and charge. The former is dominated by
large $J$ where the flux is typically treated classically and then one
considers the quantum fluctuations perturbatively. We are interested
in the opposite charging limit, where both capacitances are small and
we are in the quantum regime at the outset. To treat this case we will
proceed in two steps: first take the limit of small $C_{\text{m}}$ and
then small $C_{\text{g}}$.

\textit{Small $C_{\text{m}}$ limit}: first, we add a bias voltage $\bar q$ to each matter wire so that the kinetic term becomes
\begin{align}
	\begin{split}
		\frac{1}{2C_{\text{m}}}\;(q_n - {\bar q})^2
		\;.
	\end{split}
\end{align}
If the bias is close to $\bar q = N+1/2$ ($N$ is an integer) such that
two quantized states $q_n=N$ and $q_n=N+1$ are close in energy, then
the matter wires become two-level systems, because the small
capacitance penalizes all other charge states. For our purposes we
consider a gate bias very close to the half integer point. At this
point the operators $e^{\pm\i\, \phi_n}$ increase or decrease the
charge value and we can replace them by ordinary spin raising/lowering
operators $e^{\pm\i\, \phi_n}\to \mu^{\pm}_n$, where
$\mu^{\pm}_n = \mx_n\pm \i\my_n$.~\cite{Bouchiat_1998}. The
Hamiltonian~(\ref{eq:JJ-J}) in this limit becomes
\begin{align}
	\begin{split}
		H_J = -J\sum_{n,i} W_{ni}\;
		\left(\mp_n\;e^{+\i\theta_i}+\mm_n\;e^{-\i\theta_i}
		\right)\;.
	\end{split}
\end{align}
The $\mu$ spins can be integrated out exactly by diagonalizing, for
each $n$, a $2\times 2$ spin-1/2 Hamiltonian (treating the $\theta_i$'s
as slow fields). Keeping only the lowest energy terms, the result is
an effective potential as a function of $\theta_i$'s only:
\begin{align}
	\begin{split}
		H_J^{\text{eff}}=-|J|
		\sum_{n}\left[\sum_{i,j}W_{ni}W_{nj}\cos(\theta_i-\theta_j)\right]^{1/2}.
	\end{split}
\label{eq:Heff}
\end{align}
Notice that this Hamiltonian still has the global $U(1)$ symmetry as well as the discrete $\mathbb{Z}_2$ symmetry, which we can
write as:
\begin{align}
	\begin{split}
		\theta_i\to \theta_i + \frac{\pi}{2}(1-\sz_i),
		\;\;
		\sz_i=\pm 1, \;\text{if}\; \sz_1\,\sz_2\,\sz_3\,\sz_4 = +1\,.
	\end{split}
\end{align}

\textit{Small $C_{\text{g}}$ limit}: now we will follow a similar
procedure with the gauge wires and add a bias $\bar Q = M+1/2$ ($M$ is
an integer) to all $Q_i$. This restricts the charge on each gauge wire
to two values.

Mathematically, however, we take a different approach. Rather than
replacing the flux operators by Pauli matrices immediately, we will
first expand Eq.~(\ref{eq:Heff}) in a Fourier series, keeping only the
terms $e^{\pm\i\, \theta_i}$, i.e., those that change the charge from
one to zero or vice versa on each wire. It is straightforward to check
that the only terms that appear in the Fourier expansion have the form
of $\mathcal{A}_s$ in Eq.~(\ref{eq.A_as_pm}) but with each spin
operator $\sigma_i^{\pm}$ standing for $e^{\pm\i\, \theta_i}$.

We note that this procedure realizes the star term $\mathcal{A}_s$,
but we do not generate the plaquette term $B_p$. It remains a problem
for future work to systematically study different sectors with given
eigenvalues of $B_p$ to determine whether the ground states of
Hamiltonian Eq.~\eqref{eq:Hamiltonian} with $\lambda_B=0$ remain those
in the sector with $B_p=+1$ for all $p$, which were justified in our
Quantum Monte Carlo studies by the presence of the sufficiently large
$\lambda_B$ coupling in the Hamiltonian.

\section{Conclusions}

In this work, we studied a spin-$1/2$ lattice model -- the
$U(1)$ toric code -- that is invariant under both a global
$U(1)$ symmetry and local $\mathbb Z_2$ gauge transformations. 
We presented evidence that the system is gapped and the $U(1)$ global symmetry is not spontaneously broken.  
The exponential decay of spin-spin correlators support the claim that the system is a gapped spin liquid.
We found topologically degenerate ground states, labeled by non-contractible string operators. 

The model displays quite distinct topological degeneracies that depend on the tilt of the lattice that is wrapped around the torus, a form of UV/IR mixing unlike those encountered, for example, in fractonic models. 
The number of degenerate ground states is also puzzling. 
It is difficult to explain the three-fold topological degeneracy for the $45^\circ$-tilt
compactification as coming from Abelian topological order if the $U(1)$ toric code is described by a doubled theory (for example, the usual toric code which is described by a doubled Chern-Simons theory). 

One logical possibility is that the enrichment of the $Z_2$ toric code by the global $U(1)$ symmetry may
turn the topological order non-Abelian. We presented an
heuristic argument aimed at relating the three-fold topological degeneracy for the
$45^\circ$-tilt compactification to the two-fold topological degeneracy for the
$0^\circ$-tilt compactification based on a mapping of posited logical
operators that switch between topological ground states in both
geometries.

Finally, we presented a physical realization of the $U(1)$-symmetric
star terms in the Hamiltonian in a system of superconducting quantum
wires coupled by Josephson junctions at their crossings. We believe
that the possibility that the model may be realizable with physical
Hamiltonians should further motivate future theoretical studies of the unusual topological properties of the $U(1)$ toric
code.

\begin{acknowledgments}
  We thank Anders Sandvik for enlightening discussions and valuable
  comments. The work of K.-H.W., A.K., G.D., and C.C. is partially
  supported by the DOE Grant DE-FG02-06ER46316 (work on designing
  topological phases of matter and foundations of combinatorial gauge
  theory) and by the NSF Grant DMR- 1906325 (work on interfaced
  topological states in superconducting wires).
\end{acknowledgments}

%\newpage 
\appendix
	
\section{Hilbert space fragmentation} \label{App.Frag}

A symmetry operator $\hat{O}$, one that commutes with the Hamiltonian, block-diagonalizes the Hamiltonian into symmetry sectors associated with quantum numbers, i.e., eigenvalues of $\hat{O}$.

Sometimes, even after resolving all symmetries of the model, a symmetry block might appear to be further block-diagonalized into smaller blocks that are not associated with any obvious symmetry operators. This phenomenon is called Hilbert space fragmentation and the corresponding (quantum) dynamically disconnected blocks are called Hilbert space fragments or Krylov subsectors \cite{sala2020ergodicity,khemani2020localization,moudgalya2022hilbert}. In the thermodynamic limit, the number of Krylov subsectors originating from Hilbert space fragmentation scales much faster (proportional to the number of degrees of freedom) than the number of symmetry sectors from any type of symmetry (unless it is a local gauge symmetry) \cite{moudgalya2022hilbert}. The corresponding conserved operators (which can be constructed, e.g., by writing down a projector on the observed block in the Hamiltonian) are highly non-local and non-trivial.

\begin{figure}[ht!]
	\includegraphics[width=\linewidth]{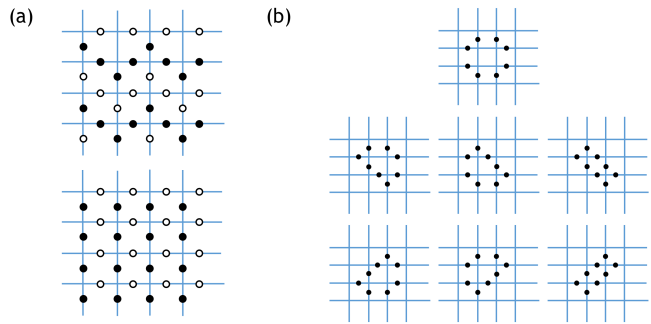}
	\caption{(a) Two states from the same symmetry sector, but from different Krylov subsectors: an inert state with no flippable stars (top), and a state where every star is flippable (bottom) (b) Basis states of a 7-dimensional Krylov subsector. Black dots denote spin-downs, white dots (or the absence of a dot) denote spin-ups.}
	\label{figfrag1}
\end{figure}

Hilbert space fragmentation typically arises in models with local kinetic constraints. Such constraints are present in the $U(1)$ toric code. As an example, consider the two states in Fig.~\ref{figfrag1}. They belong to the same symmetry sector. Namely, they belong to the block with quantum numbers: $M_z = 0$, $W_x = W_y = +1$, $B_p = +1$ for every plaquette $p$. However, the state at the top is completely inert, since none of the stars are flippable, while in the state at the bottom every star is flippable. These states are dynamically disconnected and belong to different Krylov subsectors. In fact, the inert state comprises its own 1-dimensional Krylov subsector. Another example of a 7-dimensional Krylov subsector is shown in Fig.~\ref{figfrag1}(b). If we translate this pattern of spins in space, we will obtain a different Krylov subsector with the same quantum numbers.

Hilbert space fragmentation in a model that straightforwardly maps to the topological $(+,+)$ sector of the $U(1)$ toric code has been studied in works~\cite{yoshinaga2022emergence, hart2022hilbert}. Dynamics of domain walls in the same setting has been studied in~\cite{balducci2022localization,balducci2022interface}.

\begin{figure}[ht!]
	\includegraphics[width=\linewidth]{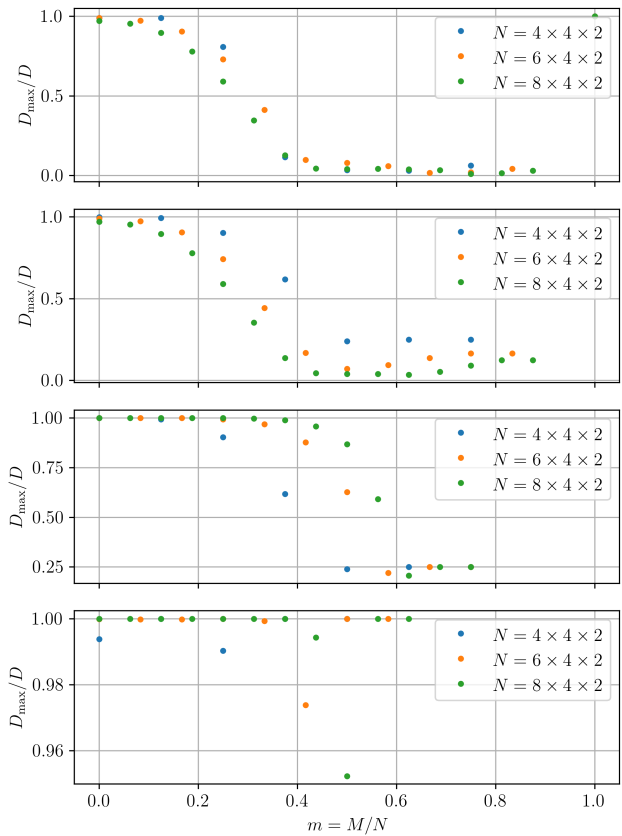}
	\caption{Exact enumeration study of the fragmentation fraction defined as $D_{max}/D$, where $D_\text{max}$ is the size of the largest fragment, and $D$ is the size of the sector. From top to bottom are the $(+,+)$, $(+,-)$, $(-,+)$ and $(-,-)$ topological sectors, respectively.}
	\label{figfrag2}
\end{figure}

Using explicit enumeration of states, we study Hilbert space fragmentation in systems of size $N = 4 \times 4 \times 2, 6 \times 4 \times 2$ and $8 \times 4 \times 2$ (the first two numbers are the numbers of stars in $x$- and $y$-directions, the last factor of 2 corresponds to the two spins in a unit cell). We concentrate on the symmetry sector with all $B_p=+1$. We define $D_\text{max}$ as the dimensionality of the largest Krylov subsector and $D$ as the dimensionality of the corresponding symmetry sector. The ratio of $D_\text{max}/D$ for different magnetization and topological symmetry sectors is presented in Fig.~\ref{figfrag2}.

One can see that in the vicinity of zero magnetization, the largest Krylov subsector completely dominates the Hilbert space of its symmetry sector. We conjecture that this behavior carries over to the thermodynamic limit. In addition, by a simple combinatorial argument, it is apparent that the largest symmetry sectors are with $M_z = 0$. Since larger random matrices have a larger spread of eigenvalues than smaller random matrices, we can safely assume that the ground state of each of the four topological symmetry sectors belongs to its largest Krylov subsector in the $M_z = 0$ sector. Work \cite{hart2022hilbert} proves this with even more rigor, although their results translate directly only to our $(+,+)$ topological sector. In addition, we confirm the results of \cite{hart2022hilbert} by observing the same ``dynamical freezing transition", which can be seen as a sharp drop of the $D_{max}/D$ value at intermediate values of magnetization [except for the $(-,-)$ sector; however, this might be a finite size effect].

\section{Generalized sweeping cluster update algorithm} \label{App.SCA}

The model we consider does not have a sign problem. Therefore, a QMC simulation is possible. We employ the SSE QMC with a modification of the sweeping cluster update algorithm, previously used to study the quantum dimer model~\cite{yan2019sweeping,MelkoRing2005}. 
	
\begin{figure}[ht!]
		\includegraphics[width=\linewidth]{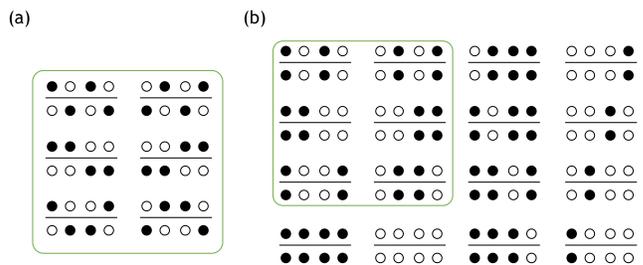}
		\caption{(a) 6 allowed off-diagonal vertices from the Hamiltonian. Black (white) dots denote spin-up (spin-down). The four dots below (above) the line denote the classical configuration of the four spins on a star before (after) the application of a Hamiltonian term. (b) 16 additional allowed diagonal vertices after adding a constant to the Hamiltonian. The 12 vertices marked by green frames are the flippable stars.}
		\label{fig1}
	\end{figure}

	As it is typically done in the SSE simulations, the partition function is expanded in a series of powers of the Hamiltonian $H$. Terms from this series are sampled as classical configurations, where slices of ``imaginary time" contain ``vertices" that take one local classical configuration to another one, according to one of the terms in the Hamiltonian. For a comprehensive review of the SSE, see \cite{sandvik2010computational}. In the $U(1)$ toric code, there are only 6 allowed off-diagonal vertices with equal weight, as shown in Fig.~\ref{fig1}(a). In order to perform the simulation, all diagonal terms have to be non-zero. To achieve this, we add a constant to the Hamiltonian, which then allows additional 16 diagonal vertices as shown in Fig.~\ref{fig1}(b). The constant is chosen such that all the vertices have equal weight. Vertices with 2-up-2-down configurations are called ``flippable stars", as marked by green frames in Fig.~\ref{fig1}. In the following, we set the weight of all vertices to 1. 
	
	Each Monte Carlo step consists of two parts. First, the standard diagonal update is performed. If an ``imaginary time" slice is empty, then a diagonal operator on a random star might be inserted; if an ``imaginary time" already contains a diagonal operator, then it might be removed. The probabilities of these two processes are
	\begin{align}
		\begin{split}
			P_{\textrm{insert}} &= \frac{\beta N_{s}}{(M-n)}, \\
			P_{\textrm{remove}} &= \frac{(M-n+1)}{\beta N_{s}}.
		\end{split}
	\end{align} 
	Here, $\beta = 1/T$ is the inverse temperature, $N_s=L_x \times L_y$ is the total number of stars, $M$ is the total length of the operator-string (i.e., the number of ``imaginary time" slices), and $n$ is the total number of non-identity operators present in the current operator-string. Off-diagonal operators are left untouched during this step.
	
\begin{figure}[ht!]
		\includegraphics[width=\linewidth]{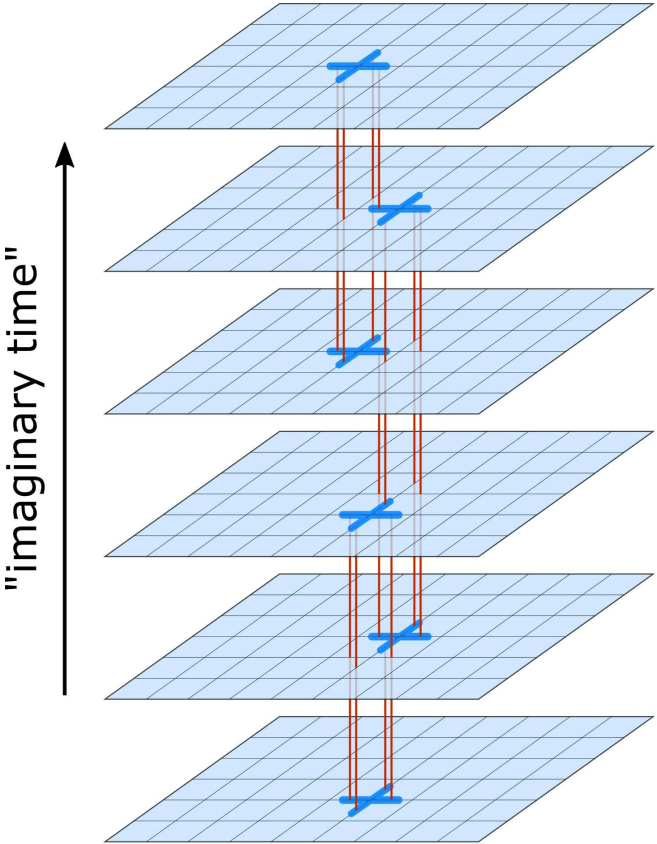}
		\caption{Example of a cluster. It starts as 4 defect lines coming out of a flippable star. The defect lines propagate upward along the ``imaginary time" direction, modifying the vertices they encounter according to specified rules. After a vertex, the number of defect lines might potentially increase/decrease. Finally, the cluster terminates at another flippable star.}
		\label{figClust}
	\end{figure}

	After the diagonal update, a sweeping cluster update is performed. The construction of a cluster starts with randomly choosing a flippable star, either diagonal or off-diagonal. This creates 4 defect lines, which propagate upward along the ``imaginary time" direction and start growing the cluster.
	We then sweep over "imaginary time" slices, and keep track of the defect lines. If a vertex is hit by one or more defect lines, we update the vertex according to specific rules and propagate the defect lines further. The number of defect lines exiting a vertex might be different from how many defect lines entered it. At some point, the cluster will converge and the defect lines will terminate at another flippable star. The pictorial representation of a cluster is shown in Fig.~\ref{figClust}
	We note that the case in which a vertex has a different numbers of entering and exiting lines has been explored in Ref.~\onlinecite{MelkoRing2005} using a different approach. Our model (as well as the dimer model) has a special property that allows the cluster to build in one directional.

\begin{figure}[ht!]
		\includegraphics[width=\linewidth]{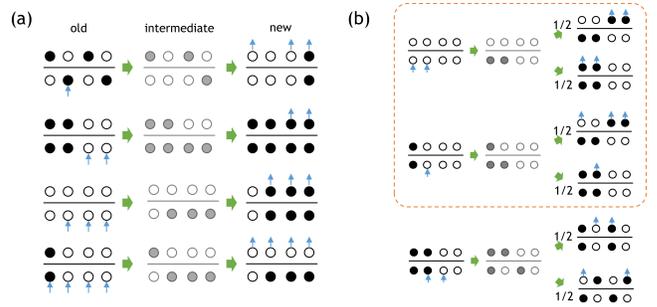}
		\caption{Examples of vertex updates. Blue arrows denote the defect lines propagating along the ``imaginary time" direction (from bottom to top). (a) If a new configuration of the bottom four spins is not flippable, there is a unique way to propagate the defect lines, such that the new vertex remains allowed. (b) If less than 4 defect lines hit a vertex and update the bottom four spins to a flippable configuration, there are two possible ways to propagate the defect lines. We choose one of them with probability 1/2. Processes marked by the orange frame are the updates from a non-flippable vertex to a flippable vertex, for which the reverse process does not have probability 1/2.}
		\label{figSSE}
	\end{figure}

	To determine the rules for the vertex updates, we proceed in the following way. Consider a vertex that is hit by some number of defect lines from below. We flip the corresponding spins and obtain an intermediate vertex configuration, as shown in Fig.~\ref{figSSE}.
	At this point, there are four possibilities: 
	\begin{enumerate}
		\item If the configuration of the bottom four spins of an intermediate vertex is non-flippable, then there is only one way to propagate the defect lines through the vertex. Some examples are shown in Fig.~\ref{figSSE}(a).
		\item If the defect lines flip the bottom four spins into a flippable configuration, and there are less than 4 lines hitting the vertex, then there are two possible ways to propagate the defect lines. The resulting vertex can be either diagonal or off-diagonal, with probability 1/2 each. Examples are shown in Fig.~\ref{figSSE}(b).
		\item If 4 defect lines hit a flippable vertex and the total number of defect lines is more than 4, then we propagate all the lines and flip the entire vertex.
		\item If 4 defect lines hit a flippable vertex and the total number of defect lines is exactly 4, then we terminate the cluster.
	\end{enumerate}

	Note that the update from a non-flippable vertex to a flippable vertex has probability 1/2, while the reverse process does not. Thus, to satisfy the detail balance, we have to keep track of how many times these processes occur. Consider an operator-string $A$ that updates to an operator-string $B$ through a flip of a cluster constructed using the aforementioned procedure. During this update, the number of non-flippable vertices that turned into flippable ones is $N_{n \rightarrow f}$, while the number of flippable vertices that turned into non-flippable ones is $N_{f \rightarrow n}$. To satisfy detailed balance, the acceptance probability for the process $A \rightarrow B$ (through the flip of the constructed cluster) should then be
	\begin{align}
		\begin{split}
			P(A \rightarrow B) = \frac{N_v(A)}{N_v(B)}\left(\frac{1}{2}\right)^{N_{f\rightarrow n} - N_{n \rightarrow f}},
		\end{split}
	\end{align}
	where $N_v$ is the number of flippable vertices present in the corresponding operator-string.
	
\begin{figure}[ht!]
		\includegraphics[width=\linewidth]{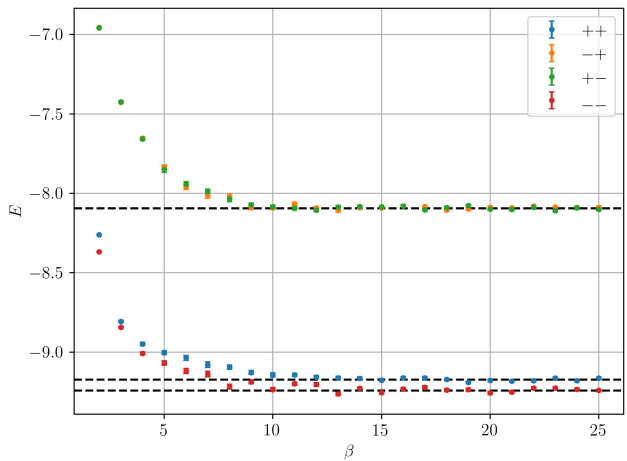}
		\caption{Ground state energy in four topological sectors of a system of size $4\times4$ stars. (dots) obtained from the SSE QMC with the generalized sweeping cluster update algorithm; (dashed lines) obtained from the ED calculation.}
		\label{figQMCBench}
	\end{figure}

	It is worth pointing out that the update procedure we described conserves all quantum numbers, $\{M_z,W_x,W_y,\{B_p\}\}$. Even more than that, the update procedure cannot take the system out of a specific Hilbert space fragment (Krylov subsector). In this way, by choosing an appropriate initial state, we are able to target a Hilbert space fragment that is dynamically connected to that initial state. As we explain in Appendix~\ref{App.Frag}, in the vicinity of zero magnetization, the Hilbert space of symmetry sectors is dominated by a single large fragment, which will most probably contain the lowest energy eigenstate.
	We benchmark our QMC results by comparing them to the results obtained from an ED calculation of a small system size. The results are presented in Fig.~\ref{figQMCBench}.
	
\section{Energies in different magnetization sectors} \label{App.Mag}

For a system with size $ L_x \times L_y$ on a square lattice, there
are $2^{L_xL_y+1}$ sectors associated to the $ \mathbb Z_2 $ plaquette
operators, $ (2L_xL_y+1) $ sectors associated to the global $U(1)$
symmetry, and $4$ sectors associated to the string operators
$ W_{x,y} $. However, not all of these sectors have states (i.e., they
are of non-zero dimension.)
	
For example, in a system with $L$ even, the sector with $\{B_p=+1\}$
and $ (W_x,W_y)=(+,+)$ is incompatible with odd magnetization
$M_z=\pm1, \pm3, \pm5, \pm7, \pm9...$, meaning that such sectors have
vanishing dimension, or equivalently, they are
non-existent. Conversely, in a system with $L$ odd, all sectors with
$\{B_p=+1\} $ and $ (W_x,W_y) = (+,+) $ can only have odd
magnetization, and consequently all even magnetization sectors
vanish.
	
\begin{figure}[h]
  \includegraphics[width=\linewidth]{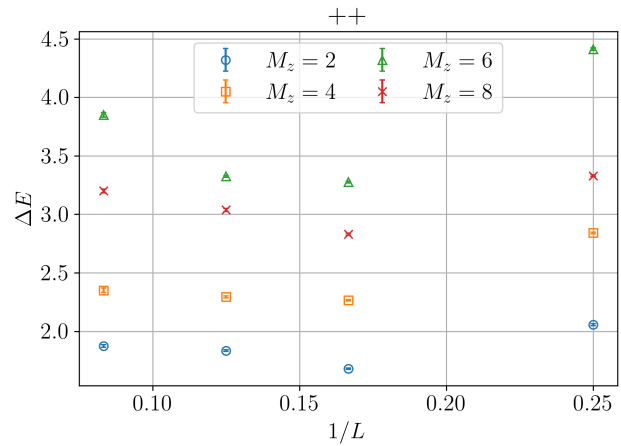}
  \caption{Energy differences between the lowest states with total magnetization $M_z=2,4,6,8$ and the lowest state with magnetization $M_z=0$, within the $(+,+)$ sector, as a function of inverse system size $1/L$. The system has $45^\circ$-tilt compactification.}
  \label{Fig.45ppgap}
\end{figure}

In Fig.~\ref{Fig.45ppgap}, we show the difference in energy between
the lowest states in each magnetization sector and the lowest state
with $M_z=0$, within the topological sector $(+,+)$. The results
indicate that the state with $M_z=0$ has the lowest energy.

\section{Spin-spin correlation functions in different sectors for $0^\circ$ and $45^\circ$ compactifications} \label{App.Corr}

The spin-spin correlation functions in Eq.~\eqref{eq.Corr} were
presented for the $(-,-)$ sector and for $0^\circ$-tilt
compactification in Fig. \ref{Fig.0degCorr}.

\begin{figure}[h]
  \includegraphics[width=\linewidth]{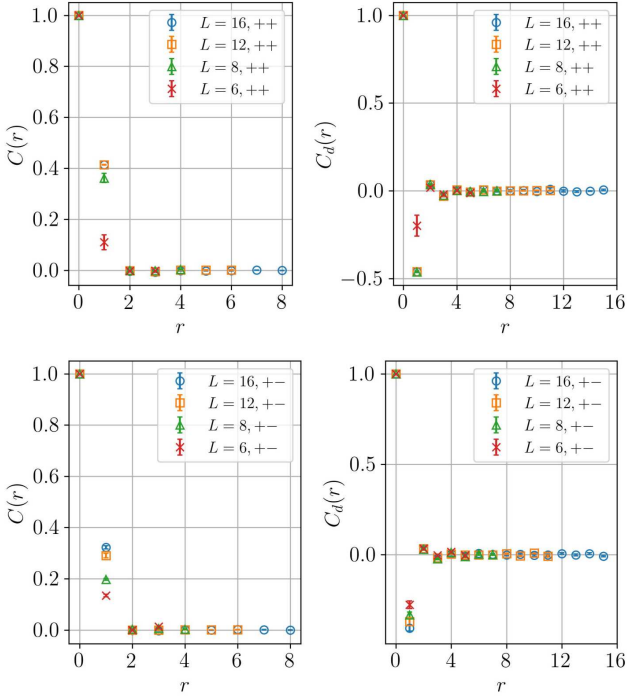}
  \caption{Spin-spin correlation functions in sectors $(+,+)$ (upper two figures) and $(+,-)$ (lower two figures) for $0^\circ$-tilt compactification. The correlations decay to zero for distances of the order of two lattice sites.}
  \label{Fig.Corr_app}
\end{figure}

We show in Fig.~\ref{Fig.Corr_app} the QMC results for spin-spin
correlations obtained in the other topological sectors, $(+,+)$ and
$(+,-)$, for $0^\circ$-tilt compactification. [The results for the
sectors $(+,-)$ and $(-,+)$ are identical by symmetry.] The data shows
a rapid decay of correlations in all topological sectors.

\begin{figure}[h]
	\includegraphics[width=\linewidth]{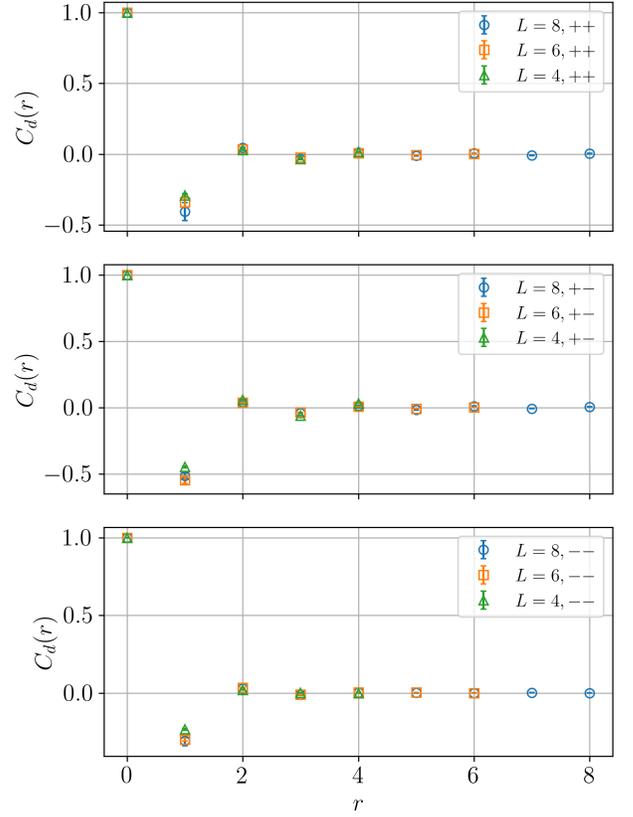}
	\caption{Spin-spin correlation functions in the topological sectors $(+,+)$, $ (+,-)/(-,+) $  and $(-,-)$  for the $45^\circ$-tilt setup. All the results shows a fast decaying within the order of two lattice sites for all system sizes $ L $.}
	\label{fig.Corr45_app}
\end{figure}

In Fig.~\ref{fig.Corr45_app}, we present the spin-spin correlation
functions for $45^\circ$-tilt compactification, for all topological
sectors. We again observe rapidly decaying spatial correlations,
consistent with the absence of long range magnetic order.

In the $45^\circ$-tilt compactification scheme we also observe that
the star-star correlators
$ \left<\mathcal{A}_{s}\;\mathcal{A}_{s'}\right>$ present a staggered
pattern, similar to what we observed for the $ 0^\circ$-tilt case and
showed in Fig. \ref{Fig.0degCorr}. This provides evidence that within
each topological sector we have a two-fold degeneracy associated to
spontaneous translational symmetry breaking. Accounting for this
additional factor of 2, the total ground state space is 6-fold
degenerate. (Note that this additional degeneracy is not topological
and can be lifted by local perturbations.)

\section{Odd system sizes results for $0^{\circ}$ and $45^{\circ}$ compactifications} \label{App.odd}

\begin{figure}[h]
  \includegraphics[width=\linewidth]{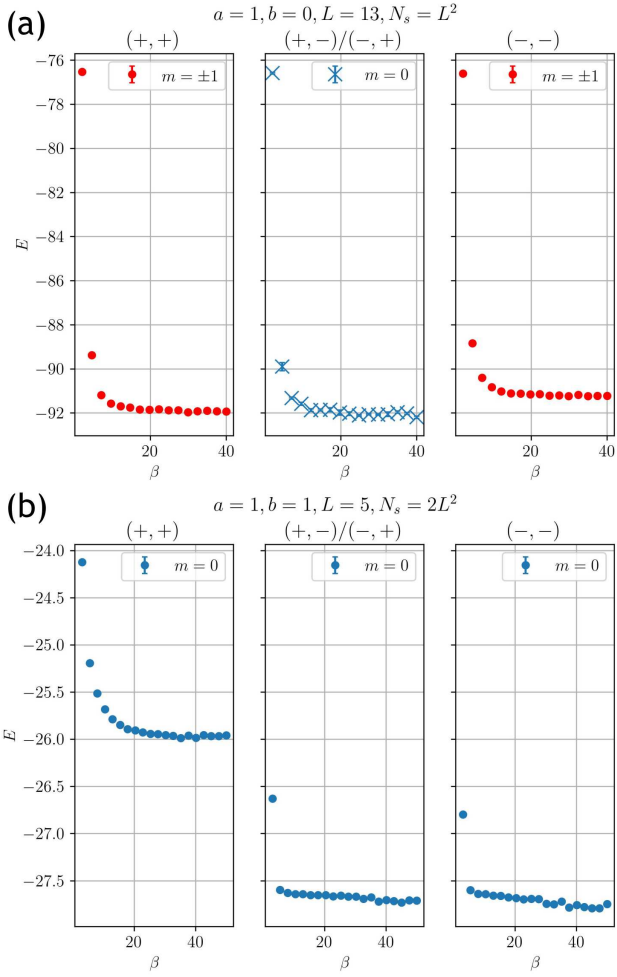}
  \caption{QMC results of energies in 4 topological sectors for (a)
    $0^{\circ}$-tilt with system size $L = 13$ where only the sectors
    (+,-) and (-,+) are compatible with zero magnetization. (b)
    $45^{\circ}$-tilt with system size $L=5$, where the TGSD remains
    three as the even size system, but the degenerate sectors are
    (+,-) (-,+) and (-,-). $N_s$ denotes the total number of stars on
    the lattice.}
  \label{fig.oddsz}
\end{figure}

In Fig.~\ref{fig.oddsz}, we show the QMC results of energy in 4
different topological sectors with odd system size $L$ on both the
$0^{\circ}$ and $45^{\circ}$ compactifications.

With odd system size $L$, as shown in Fig.~\ref{fig.oddsz}(a) for the
case of $0^{\circ}$-tilt, we find that sectors $(+,+)$ and $(-,-)$ are
not compatible with even magnetization. Therefore, the lowest energy
levels in these sectors has $M_z = \pm 1$, not $M_z=0$. These states
can be split by a local longitudinal field perturbation term in the
Hamiltonian. On the other hand, sectors with $(+,-)$ and $(-,+)$ are
compatible with even magnetization, and the lowest energy states have
$M_z=0$. Based on the above observation, we believe that in case the
topological order is present in the system, the TGSD has to come from
the $(+,-)$ and $(-,+)$ sectors.
	
In the case of $45^{\circ}$-tilt, all four topological sectors are
compatible with even magnetization and the lowest energy states have
$M_z=0$. As shown in Fig.~\ref{fig.oddsz}(b), we find that states from
$(+,-)$, $(-,+)$ and $(-,-)$ become degenerate at low temperatures
($\beta = 1/T$), indicating a three-fold TGSD. Compared to the even
$L$ case, the TGSD stays the same, but instead of the $(+,+)$ sector,
the reference sector is now $(-,-)$.
	
\section{ED results for other compactifications} \label{App.GenCptfy}	

\begin{figure}[h]
  \includegraphics[width=\linewidth]{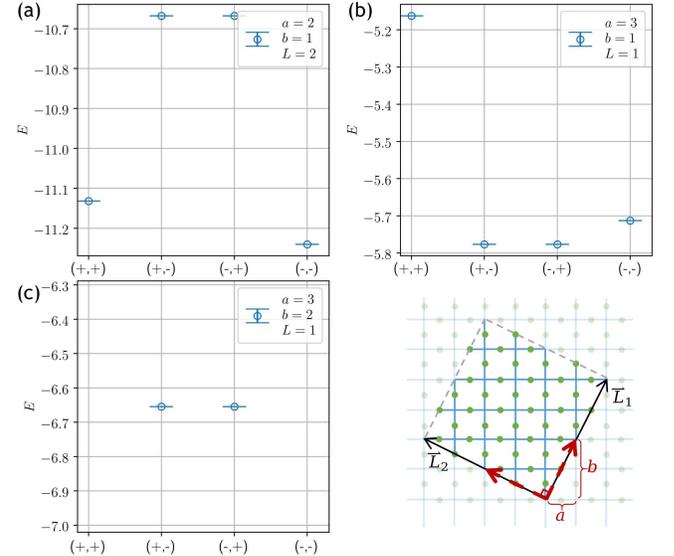}
  \caption{ED results for the energy of the lowest states in each
  topological sectors with $m=0$ for different compactifications. (a)
  $a=2 , b=1, L=2$. (b) $a=3, b=1, L=1$ (c) $a=3, b=2, L=1$.}
  \label{fig.Comp}
\end{figure}

For other compactification schemes, we study the ground state energy
in different topological sectors using ED (see
Fig.\,\ref{fig.Comp}). The total number of stars is
$N_s = (a^2 + b^2)L^2$, and the total number of spins is $N =
2N_s$. Here, we show the results only for zero magnetization, as we
find that the states with $M_z = 0$ have lower energy than other
states non-zero magnetization states.
	
Due to size limitations to the computations, it is only possible to
study compactification schemes with small values of $a$ and $b$.  Here
we show results for three different cases: a) $a=2,b=1$ and $L=2$
($N_s = 20$); b) $a=3,b=1$ and $L=1$ ($N_s = 10$); and c) $a=3,b=2$
and $L=1$ ($N_s = 13$). In case (a), we find energy features similar
to the $0^{\circ}$-tilt case with an even system size $L$. The lowest
energy states of sectors $(+,+)$ and $(-,-)$ are well separated from
the lowest energy states of $(+,-)$ and $(-,+)$ sectors, and are,
possibly, the two topologically degenerate ground states, with the
small energy difference between them being a finite size
effect. Likewise, for (b), we find that the energies behave similarly
to the $45^{\circ}$-tilt case. The $(+,-)$, $(-,+)$ and $(-,-)$
sectors contain the 3 topologically degenerate ground states, again,
separated by a small finite size gap and well separated from the
$(+,+)$ sector. In case (c), sectors $(+,+)$ and $(-,-)$ are
incompatible with even magnetization, and states from the $(+,-)$,
$(-,+)$ sectors are degenerate, similarly to the $0^{\circ}$-tilt case
with odd system size $L$.
	
Based on the above observations, we find that the TGSD is always
either 2 or 3, depending on the compactification scheme. The results
can be summarized in the following way: if $(a^2 + b^2)$ is even,
$\mathrm{TGSD}=3$; if $(a^2 + b^2)$ is odd, $\mathrm{TGSD}=2$.  In the
case of $\mathrm{TGSD}=3$, if $L$ is even, the ground states belong to
the $(+,+),(+,-),(-,+)$ sectors, while if $L$ is odd, they belong to
the $(-,-),(+,-),(-,+)$ sectors.  In the case of $\mathrm{TGSD}=2$,
for even $L$, the ground states lie in the $(+,+)$ and $(-,-)$
sectors, while for odd $L$, the $(+,+)$ and $(-,-)$ sectors are not
compatible with zero magnetization and the ground states lie in
sectors $(+,-),(-,+)$.
	
\bibliographystyle{apsrev4-1}
\bibliography{refs.bib}

\end{document}